%% file: main.tex
\UseRawInputEncoding
\pdfoutput=1
\documentclass{article}
\usepackage{iclr2025_conference,times}
\usepackage{hyperref}
\usepackage{url}
\usepackage{xspace}
\usepackage{graphicx}
\usepackage{tcolorbox}
\usepackage{booktabs}
\usepackage{multirow}
\usepackage{enumitem}
\usepackage{amsmath}
\usepackage{amsfonts}

\title{IntrinsicVoice: Empowering LLMs with Intrinsic Real-time Voice Interaction Abilities}

\author{
    \textbf{Xin Zhang$^{1}$\thanks{Work done during internship in Tongyi Speech Laboratory, Alibaba Group}},
    ~\textbf{Xiang Lyu$^{2}$},
    ~\textbf{Zhihao Du$^{2}$},
    ~\textbf{Qian Chen$^{2}$},
    ~\textbf{Dong Zhang}$^{1}$,\\
    ~\textbf{Hangrui Hu$^{2}$},
    ~\textbf{Chaohong Tan$^{2}$},
    ~\textbf{Tianyu Zhao$^{2}$},
    ~\textbf{Yuxuan Wang$^{2}$},
    ~\textbf{Bin Zhang$^{2}$},\\
    ~\textbf{Heng Lu$^{2}$\thanks{Corresponding author}},
    ~\textbf{Yaqian Zhou$^{1}$}\footnotemark[\value{footnote}],
    ~\textbf{Xipeng Qiu$^{1}$}\footnotemark[\value{footnote}] \\
    $^{1}$School of Computer Science, Fudan University\\
    $^{2}$Tongyi Speech Laboratory, Alibaba Group\\
    {\tt 	\{xin\_zhang22\}@m.fudan.edu.cn} \\
    {\tt 	\{zhouyaqian,xpqiu\}@fudan.edu.cn} \\
    {\tt \{h.lu\}@alibaba-inc.com}
}

\newcommand{\method}{IntrinsicVoice\xspace}

\iclrfinalcopy

\begin{document}

\maketitle

\begin{abstract}
\input{Sections/00_abstract}
\end{abstract}

\section{Introduction}
\input{Sections/10_introduction}

\section{Related Work}
\input{Sections/20_related_work}

\section{\method}
\input{Sections/30_method}

\section{Experiments}
\input{Sections/40_experiments}

\section{Case Study}
\input{Sections/50_case_study}

\section{Conclusion}
\input{Sections/60_conclusion}

\bibliography{main}

\appendix
\section{Appendix}
\input{Sections/70_appendix}

\end{document}

%% file: Sections/00_abstract.tex
Current methods of building LLMs with voice interaction capabilities rely heavily on explicit text autoregressive generation before or during speech response generation to maintain content quality, which unfortunately brings computational overhead and increases latency in multi-turn interactions. To address this, we introduce \method, an LLM designed with intrinsic real-time voice interaction capabilities. \method aims to facilitate the transfer of textual capabilities of pre-trained LLMs to the speech modality by mitigating the modality gap between text and speech.
Our novelty architecture, GroupFormer, can reduce speech sequences to lengths comparable to text sequences while generating high-quality audio, significantly reducing the length difference between speech and text, speeding up inference, and alleviating long-text modeling issues. Additionally, we construct a multi-turn speech-to-speech dialogue dataset named \method-500k which includes nearly 500k turns of speech-to-speech dialogues, and a cross-modality training strategy to enhance the semantic alignment between speech and text. Experimental results demonstrate that \method can generate high-quality speech response with latency lower than 100ms in multi-turn dialogue scenarios. Demos are available at \href{https://instrinsicvoice.github.io/}{https://instrinsicvoice.github.io/}.

%% file: Sections/10_introduction.tex
Large language models (LLMs) \citep{yang2024qwen2, dubey2024llama, openai2023gpt} and multimodal large language models (MLLMs) \citep{tang2023salmonn,chu2024qwen2,liu2024visual} have exhibited exceptional performance across a variety of natural language processing tasks and multimodal comprehension tasks, allowing them to become powerful solvers for general tasks. These models assist humans in solving diverse problems through text-based multi-turn dialogue interactions. However, this single mode of interaction limits the applicability of LLMs, hindering their deeper integration into everyday life. Recently, GPT-4o \citep{openai2024gpt4o} has showcased real-time, efficient voice interaction capabilities, responding to users' speech commands with low-latency and generating high-quality speech responses. This enhancement significantly improves user experience and broadens the application scenarios for LLMs. Yet, the open-source community still lacks successful explorations in constructing such LLMs with multi-turn efficient and low-latency voice interaction capability. How to empower LLMs with low-latency, high-quality voice interaction abilities in multi-turn dialogue scenarios remains a critical challenge to be solved.

Traditionally, LLMs with voice interaction capability mainly adopt a cascading paradigm \citep{huang2024audiogpt}, i.e. the LLM is connected with an automatic speech recognition (ASR) model and a text-to-speech (TTS) model in tandem. The ASR model transcribes the user's speech instruction into text and the TTS model synthesizes response from LLM into speech. However, this cascaded approach suffers from error accumulation and the loss of paralinguistic information such as emotion and prosody, and the overall system tends to have higher latency. 

\citet{zhang2023speechgpt} proposed SpeechGPT, an end-to-end method that discretizes speech waveforms into discrete tokens and extends LLM's vocabulary to support speech input and output. Theoretically, models built in this approach can generate speech responses directly from speech instructions. However, the modality gap between speech and text poses significant challenges in transferring textual reasoning abilities to the speech domain. Existing work circumvents the modality alignment challenge by introducing text autoregressive generation before or when generating speech responses. This approach allows text to guide the generation of speech responses, thereby enhancing the quality of response content. \citet{zhang2023speechgpt} proposed a prompting approach named Chain-of-Modality (CoM), which generates intermediate text before producing a speech response to improve the quality of the response content, but at the cost of increased response latency. 
\citet{mitsui2024pslm} introduced PSLM, which generates text and speech tokens in parallel through multiple input-output sequences to reduce latency while maintaining the quality of speech response content. Nevertheless, the need for the ASR model on the input side introduces additional latency. 
\citet{xie2024mini} and \citet{fang2024llama} integrate the Whisper \citep{radford2023robust} encoder as their speech encoder addressing the issue of requiring text input. During the generation phase, they simultaneously generate text tokens and discrete speech tokens extracted by either SNAC \citep{Siuzdak2024snac} or HuBERT \citep{hsu2021hubert}. Yet, since these approaches employ different speech representations for input and output, in multi-turn dialogue scenarios, the Whisper encoder must re-encode the generated speech responses before the next dialogue turn, which introduces additional computational overhead and increases latency.   

In this paper, we propose \textbf{\method}, a large language model with intrinsic real-time voice interaction ability, capable of generating speech responses directly from speech instructions. Distinct from previous work, to avoid the additional computational overhead and latency in multi-turn dialogue scenarios, \method eliminates the need for explicit text autoregressive generation either before or during the generation of speech responses. It aims to achieve high-quality speech response by \textbf{bridging the modality gap between speech and text} and facilitating the transfer of textual capabilities to the speech domain. This is accomplished by \textbf{narrowing the sequence length difference} between text and speech tokens and \textbf{enhancing semantic alignment} between the two modalities.

Typically, speech token sequences span several times the length of text token sequences, leading to a substantial disparity in semantic information density between the two modalities. This difference significantly affects how models compute attention when processing these sequences, making the capabilities of textual Large Language Models transfer to the speech domain more difficult. Moreover, the excessive length of speech sequence also reduces inference speed and leads to issues with long-context modeling. To address this challenge, inspired by \citet{chen2024vall}, we partition the speech token sequence into specified-sized groups, merging all speech tokens within each group into a single frame. This approach circumvents the frame rate limitations inherent to the speech encoder, making reducing the speech frame sequence to a length comparable to that of the text token sequence possible. To maintain the generated speech quality at such meager frame rates, we propose \textbf{GroupFormer}, which augments an LLM with a smaller transformer-encoder-based Group Model. At each time step, the Group Model non-autoregressively predicts a group of speech tokens based on the hidden states output by the LLM. 


Previous work \citep{zhang2023speechgpt, rubenstein2023audiopalm, fang2024llama, xie2024mini} has primarily utilized text-speech paired data from TTS and ASR tasks to train LLMs modeling $P(speech|transcription)$ and $P(transcription|speech)$, to align the two modalities in semantic. To further enhance the semantic alignment between text and speech modalities, we propose a new training strategy that constructs multiple cross-modal tasks from a speech-to-speech dataset. By training the model on these tasks simultaneously, we aim to deepen its understanding of the semantic consistency between speech and text, thereby improving overall semantic alignment.

The contribution of this work can be summarized in the following:
\begin{itemize}[leftmargin=*]
    \item We introduce \method, a speech-text Large Language Model equipped with intrinsic multimodal understanding and generation capabilities, capable of engaging in multi-turn real-time voice interactions.
    \item We propose GroupFormer, a model that effectively reduces speech token sequences to lengths comparable to text sequences while generating high-quality audio. This approach mitigates the modality gap between speech and text, accelerates inference speed, and alleviates issues associated with long-sequence modeling.
     \item We propose a training strategy for speech-text Large Language Models to enhance the semantic alignment between them.
    
\end{itemize}

%% file: Sections/20_related_work.tex
\textbf{Multi-modal Large Language Model}
A typical approach to constructing multimodal LLMs involves utilizing a well-pretrained vision or audio encoder to obtain visual or audio embeddings, which are then aligned with the text input of LLM through a lightweight adaptor \citep{liu2024visual, chen2024ll3da, li2023blip, chu2024qwen2, tang2023salmonn}. While this method enables LLMs to develop multimodal understanding abilities, their output remains constrained to the text modality. On the other hand, works like SpeechGPT \citep{zhang2023speechgpt}, LauraGPT \citep{chen2023lauragpt}, and AnyGPT \cite{zhan2024anygpt} proposed converting continuous vision/audio representations into discrete tokens and incorporating these discrete tokens into the LLMs' vocabulary, thereby empowering LLMs with multimodal generation capabilities. Yet, how to transfer the knowledge of pretrained LLM to other modalities seamlessly, consequently obtaining better generalization and human-instruction following ability, remains an unresolved challenge.

\textbf{Audio Language Modeling} With the success of language models in natural language processing and the development of discrete audio representations for audio understanding and generation \citep{hsu2021hubert, chung2021w2v, zeghidour2021soundstream, defossez2022high, zhang2024speechtokenizer}, language models have been widely applied to various speech generation tasks. \citet{lee2021direct, zhang2023dub} utilized HuBERT units \citep{hsu2021hubert} for speech translation, while the VALL-E series \citep{wang2023neural, zhang2023speak, chen2024vall} employed audio codecs for zero-shot text-to-speech (TTS) synthesis. AudioLM \citep{borsos2023audiolm} proposed a hierarchical approach that first models semantic tokens and subsequently audio codecs from the generated semantic tokens, enabling consistent speech synthesis in a textless setting. SPEAR-TTS\citep{kharitonov2023speak} and CosyVoice \citep{du2024cosyvoice} further adopted this hierarchical method in zero-shot TTS. \citet{zhang2024speechgpt} observed that disentangled semantic modeling and perceptual modeling can enhance performance in both dimensions. The excessive length of audio token sequences is a challenge of audio language modeling, which not only results in slow inference speed but also leads to issues of long-context modeling. \citet{lee2021direct} proposed collapsing consecutive sequences of identical units into a single unit to reduce the length of audio token sequences, which led to performance improvements in speech translation tasks. However, the resulting sequence length still differs significantly from text sequences. \citet{chen2024vall} introduced a grouping strategy that partitions the speech token sequence into specified-size groups and models each group as a single frame, reducing the frame rate by integer multiples. Nevertheless, this method suffers from noticeable performance degradation at high compression rates.

\textbf{Real-Time Voice Interaction with LLMs} Early efforts to enable voice interaction with LLMs primarily employed a cascading paradigm, where ASR and TTS models were connected in tandem with the LLM. However, this approach suffers from error accumulation, loss of paralinguistic information, and increased latency. SpeechGPT \citep{zhang2023speechgpt}, AudioPalm \citep{rubenstein2023audiopalm}, Spectron \citep{nachmani2023spoken}, and SpeechGPT-Gen \citep{zhang2024speechgpt} have adopted an end-to-end method using Chain-of-Modality (CoM) prompting, which generates intermediate text before producing a speech response. While this approach improves the quality of the generated speech content, it comes at the cost of increased latency due to the text autoregressive generation before generating speech. To address this problem, PSLM \citep{mitsui2024pslm} generates text and speech in parallel, but the reliance on ASR models for input still introduces additional latency. Mini-Omni \citep{xie2024mini} and Llama-Omni \citep{fang2024llama} tackle this issue by feeding continuous speech embeddings from Whisper \citep{radford2023robust} encoder into the LLM, enhancing model's ability to understand speech instructions without requiring text input. However, the inconsistency between speech input and output introduces additional computational overhead, increasing latency in multi-turn scenarios. Moshi \citep{defossezmoshi} seeks to resolve this complication by employing an audio tokenizer that integrates both semantic and acoustic information, and by temporally aligning text and audio tokens.

%% file: Sections/30_method.tex
\begin{figure}[h]
\begin{center}
\includegraphics[width=1\textwidth]{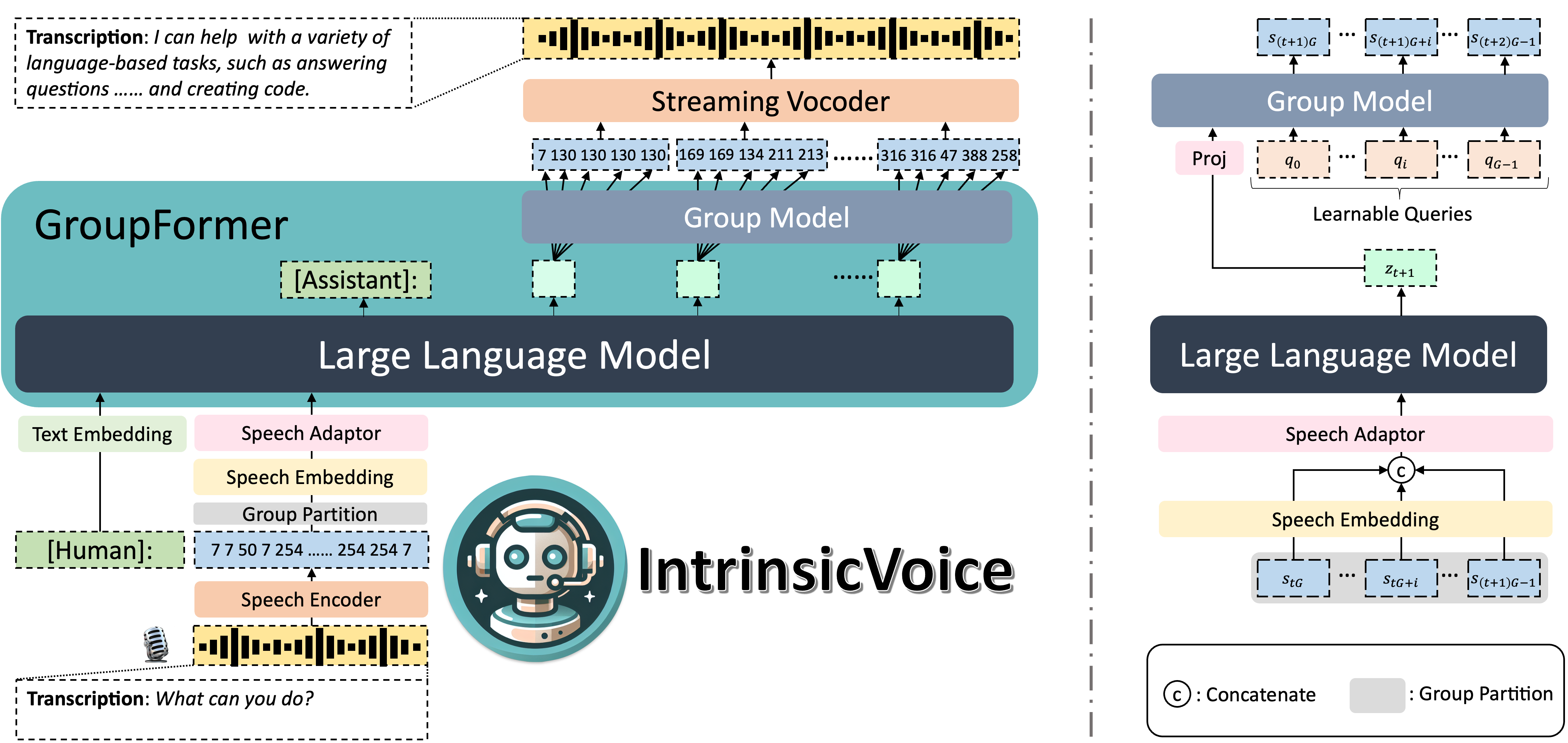}
\end{center}
\caption{\textbf{Left}: The overview of the \method model architecture. \textbf{Right}: Illustration of the proposed GroupFormer model structure.}
\label{fig:overview}
\end{figure}

In this section, we introduce the architecture of \method. As depicted in Figure\ref{fig:overview}, our model consists of the following main components: a speech encoder and a streaming vocoder for the tokenization and detokenization of speech; a speech embedding layer and a speech adaptor for mapping grouped discrete speech tokens into embeddings of LLM; and a GroupFormer for modeling speech and text sequence, which consists of a large language model and a group model. 

\subsection{Speech Tokenization}
\label{sec:31_tokenization}
\input{Sections/31_tokenization}

\subsection{Embedding Grouped Speech Tokens}
\label{sec:32_embedding_group}
\input{Sections/32_embedding_group}

\subsection{GroupFormer}
\label{sec:33_GroupFormer}
\input{Sections/33_groupformer}


\subsection{Training}
\input{Sections/35_train}

%% file: Sections/31_tokenization.tex
\textbf{Speech Tokenization} Speech tokenization involves transforming continuous speech signals into a sequence of discrete speech tokens, which enables language models to handle speech like text tokens. Semantic tokens, typically from self-supervised pre-trained models with masked language modeling as training objective, have been widely used for speech tokenization \citep{zhang2023speechgpt, rubenstein2023audiopalm, hassid2024textually} due to their strong alignment with text \citep{zhang2024speechtokenizer}. Following \citet{zhang2023speechgpt} and \citet{mitsui2024pslm}, we employ HuBERT \citep{hsu2021hubert} as speech encoder for speech tokenization, converting speech waveform $\mathbf{A}$ into speech token sequence $\mathbf{S}=[s_0, s_1, \cdots, s_{T-1}]$, where $T$ is the speech tokens length.

\textbf{Speech Detokenization} Speech detokenization refers to the process of converting discrete speech tokens $\mathbf{S}$ back into speech waveform $\mathbf{A}$. We use HiFi-GAN \citep{kong2020hifi}, a non-autoregressive neural vocoder that efficiently generates high-fidelity waveforms, for speech detokenization to reduce overall latency. Following the approach in \citet{mitsui2024pslm}, we provide partial speech tokens to generate waveform fragments. Specifically, we utilize non-causal convolution to guarantee high speech quality, and the first speech fragment can be generated once $N_{\text{offset}}=\lfloor R/2\rfloor + 1$ tokens are decoded, where $R$ denotes the receptive field of HiFi-GAN.

%% file: Sections/32_embedding_group.tex
To reduce the length of speech token sequence $\mathbf{S}$, we partition it into a grouped token sequence $\mathbf{S}^G=[\mathbf{S}_{0:G}, \mathbf{S}_{G:2G}, \cdots, \mathbf{S}_{(T-G):T}]$, where $G$ denotes group size and $\mathbf{S}_{0:G}$ stands for the group $[s_0, s_1, \cdots, s_{G-1}]$. Following \citet{chen2024vall}, we clip a few tokens from the start of the speech token sequence to ensure the token sequence length $T$ is the integer multiple of group size. 

As shown in the right part of Figure \ref{fig:overview}, a speech embedding layer transforms the discrete grouped speech token sequence $\mathbf{S}^G$ into a dense embedding sequence $[\mathbf{e}_{0:G}^S, \mathbf{e}_{G:2G}^S, \cdots, \mathbf{e}_{(T-G):T}^S]$, where $\mathbf{e}_{0:G}^S=[\mathbf{e}_0^S, \mathbf{e}_1^S, \cdots, \mathbf{e}_{G-1}^S]$. Each group token embedding is concatenated in the hidden dimension to produce the grouped embedding sequence $\mathbf{E}^{G}=[\mathbf{e}_{0}^{G}, \mathbf{e}_{1}^{G}, \cdots, \mathbf{e}_{T/G-1}^{G}]$. Subsequently, a trainable speech adaptor is used to map these grouped embeddings into the embedding space of the LLM. This speech adaptor is a single hidden layer followed by an ELU activation and a regression layer, denoted as:
\begin{align}
    \mathbf{E}^{SG}=Linear(ELU(Linear(\mathbf{E}^{S}))),
\end{align}
where $\mathbf{E}^{SG}=[\mathbf{e}_{0}^{SG}, \mathbf{e}_{1}^{SG}, \cdots, \mathbf{e}_{T/G-1}^{SG}]$.

The prompt template is illustrated in Figure \ref{fig:prompt}. The tokens \texttt{<sosp>} and \texttt{<eosp>} are used to denote the start and end of speech sequences, respectively. The token \texttt{<speech>} indicates the presence of a group of speech tokens $\mathbf{S}_{tG:(t+1)G}$. This template is tokenized into the token sequence $\mathbf{Y}=[y_0, y_1, \cdots, y_{L-1}]$, where $L$ is the length of $\mathbf{Y}$. Positions in $\mathbf{Y}$ corresponding to the \texttt{<speech>} token are filled with the respective grouped speech embeddings $\mathbf{e}^{SG}$ from $\mathbf{E}^{SG}$, while tokens in other positions are converted into dense embeddings using the pretrained LLM's text embedding layer. This results in the formation of the embedding sequence $\mathbf{E}=[\mathbf{e}_{0}, \mathbf{e}_{1}, \cdots, \mathbf{e}_{L-1}]$ that is inputted into the LLM.

\begin{figure*}[htb]
    \centering
    \begin{tcolorbox}[width=1\textwidth]
\texttt{<|im\_start|>} system

You are a helpful voice-chat assistant\texttt{<|im\_end|>}

\texttt{<|im\_start|>} user

[Text Context](\textcolor{red}{Optional})

\texttt{<sosp>}\texttt{<speech>}...\texttt{<speech>}\texttt{<eosp>}\texttt{<|im\_end|>}

\texttt{<|im\_start|>}assistant

\texttt{<sosp>}\texttt{<speech>}...\texttt{<speech>}\texttt{<eosp>}\texttt{<|im\_end|>}

\texttt{<|im\_start|>}user

\texttt{<sosp>}\texttt{<speech>}...\texttt{<speech>}\texttt{<eosp>}\texttt{<|im\_end|>}

\texttt{<|im\_start|>}assistant

\texttt{<sosp>}\texttt{<speech>}...\texttt{<speech>}\texttt{<eosp>}\texttt{<|im\_end|>}

.....
\end{tcolorbox}
    \caption{Prompt template of \method}
    \label{fig:prompt}
\end{figure*}

%% file: Sections/33_groupformer.tex


\citet{lee2021direct} observed that the "stacked" strategy, which is similar to our group strategy, achieved worse performance than the "reduced" strategy, which represents collapsing a consecutive sequence of the same units into a single unit, resulting in a sequence of unique discrete units. We attribute this to the high compression rate causing a group of speech tokens to probably contain distinct HuBERT units, which have temporal dependencies among them. The approach of using multiple linear heads to parallelly predict a group of tokens exhibits weaker context modeling capabilities, struggling to handle such temporal dependencies within the group. To mitigate this issue without significantly increasing latency, as illustrated in Figure \ref{fig:overview}, we introduced a GroupFormer, which augments the LLM with a smaller non-autoregressive transformer encoder model (dubbed "GroupModel") designed to enhance the model's ability to model sequences within groups.

Specifically, for a given sequence step $0<t<L$, the LLM maps $\mathbf{E}_{\leq t}=[\mathbf{e}_0, \mathbf{e}_1, \cdots, \mathbf{e}_{t}]$ to a global context embeddings $\mathbf{z}_{t+1}\in\mathbb{R}^d$, where $d$ denotes the dimension of the LLM's hidden states. The prediction head of the LLM maps $\mathbf{z}_{t+1}$ to logits estimates $p_{t+1}\in \mathbb{R}^{N}$, where $N$ denotes the size of the LLM's vocabulary. This vocabulary is an extension of the original pretrained LLM vocabulary, enriched with \texttt{<sosp>}, \texttt{<eosp>} and \texttt{<speech>}. The predicted token $y_{t+1}$ is then sampled from the distribution obtained by applying softmax function to $p_{t+1}$.

If the currently predicted token $y_{t+1}$ is  \texttt{<speech>}, $\mathbf{z}_{t+1}$ is transformed through a projection layer and then concatenated with a $G$ learnable queries, forming embedding sequence $\mathbf{E}^{GM}_{t+1}=[proj(\mathbf{z}_{t+1}), \mathbf{q}_{0}, \mathbf{q}_{1}, \cdots, \mathbf{q}_{G-1}]$. Taking $\mathbf{E}^{GM}_{t+1}$ as input, the GroupModel predicts a group of speech tokens in one step, denoted as $\mathbf{S}_{(t+1)G:(t+2)G}=[s_{(t+1)G}, s_{(t+1)G+1}, \cdots, s_{(t+2)G-1}]$. These tokens are subsequently processed as outlined in Section \ref{sec:32_embedding_group}, producing the input embedding $\mathbf{e}_{t+1}$ for the LLM at step $t+1$. Conversely, if $y_{t+1}$ is not  \texttt{<speech>},  the LLM’s text embedding layer directly maps $y_{t+1}$ into the LLM's input embedding $\mathbf{e}_{t+1}$ for the subsequent step.

The GroupFormer is trained with two training objectives. The first objective is to autoregressively generate $\mathbf{Y}$, which is formulated as:
\begin{align}
    \mathcal{L}_{LLM}=-\sum_{t=0}^{L}\log p(y_t|y_{<t},s_{<t'G};\theta_{LLM}),
\end{align}
where $\theta_{LLM}$ denotes the parameters of the LLM and $t'$ represents the number of \texttt{<speech>} tokens contained in $y_{<t}$. The second objective is to generate $\mathbf{S}^{G}$, expressed as:
\begin{align}
    \mathcal{L}_{G}&=-\sum_{t'=0}^{T/G-1}\log p(\mathbf{S}_{t'G:(t'+1)G}|y_{<t},s_{<t'G};\theta)\\
    &=-\sum_{t'=0}^{T/G-1}\sum_{i=t'G}^{(t'+1)G-1}\log p(s_i|y_{<t},s_{<t'G};\theta),
\end{align}
where $T$ is the length of the speech token sequence and $\theta$ represents the parameters of the GroupFormer, encompassing both the LLM and the GroupModel.

%% file: Sections/35_train.tex
Speech-to-speech data is structured as quadruples of the form (\emph{SI}, \emph{IT}, \emph{SR}, \emph{RT}), where \emph{SI} denotes speech instruction, \emph{IT} denotes instruction transcription, \emph{SR} denotes speech response, and \emph{RT} denotes response transcription. We reformulate these tuples into four distinct tasks: (1) \emph{SI}$\rightarrow$\emph{SR}, (2) \emph{SI}$\rightarrow$\emph{RT}, (3) \emph{IT}$\rightarrow$\emph{SR}, and (4) \emph{IT}$\rightarrow$\emph{RT}. By this approach, we construct multiple cross-modal tasks from a single dataset, enabling the model to better learn the semantic consistency between speech and text.

To fully leverage the speech data, the losses $\mathcal{L}_{G}$ is computed on both the input and output speech, whereas $\mathcal{L}_{LLM}$ is computed solely on the response portion. Consequently, the overall training objective $\mathcal{L}$ can be formulated as follows:
\begin{align}
    \mathcal{L} = \mathcal{L}_{LLM}+\mathcal{L}_{G}.
\end{align}

%% file: Sections/40_experiments.tex
\subsection{Experimental Setups}
\label{sec:41_setups}
\input{Sections/41_setups}

\subsection{Baselines}
\label{sec:42_baselines}
\input{Sections/42_baselines}

\subsection{Evaluation}
\label{sec:43_evaluation}
\input{Sections/43_evaluation}

\subsection{Main Results}
\label{sec:44_main_results}
\input{Sections/44_main_results}

%% file: Sections/41_setups.tex
\textbf{Datasets} 
 We utilize CosyVoice-300M-SFT model \citep{du2024cosyvoice} to synthesize 89k multi-turn speech QA pairs (comprising 345k turns in total) from the Moss-002-sft-data\footnote{https://huggingface.co/datasets/fnlp/moss-002-sft-data} \citep{sun2024moss} and CoQA\footnote{https://huggingface.co/datasets/stanfordnlp/coqa} \citep{reddy2019coqa} datasets. Additionally, we use the SQuAD\footnote{https://huggingface.co/datasets/rajpurkar/squad} dataset \citep{rajpurkar2016squad} to synthesize 87k single-turn speech QA pairs. By incorporating 54k samples from the Spoken-Alpaca-GPT\footnote{https://huggingface.co/datasets/GSQA/spoken-alpaca-gpt4} dataset, we construct our multi-turn speech-to-speech QA dataset, Intrinsic\_500k, which consists of approximately 500k QA turns. We employ around 20k hours from Multilingual Librispeech \citep{pratap2020mls} for ASR and TTS tasks to further enhance the model's speech understanding and synthesis capabilities. To maintain the LLM's textual abilities, we sample 350k samples from the Guanaco\_Belle\_Merge\_v1.0\footnote{https://huggingface.co/datasets/Chinese-Vicuna/guanaco\_belle\_merge\_v1.0} \cite{leng2023chinese-vicuna} text QA dataset for training. Datasets are detailed in Tabel \ref{tab:data}.

\input{Tables/data}

\textbf{Model Configuration} For the speech tokenizer, we utilize mhubert-base-25hz and its corresponding KMeans quantizer from \citet{hassid2024textually}, operating at 25Hz with 500 clusters. Our backbone LLM is Qwen2-7B-Instruct \citep{yang2024qwen2}, and the GroupModel is an 8-layer, 16-head, 512-hidden-size non-causal transformer encoder with learnable position embeddings. We partition the speech tokens into group sizes of 5, resulting in a 5 tokens-per-second (TPS) 
 speech sequence, closely matching the 4.2 TPS we observed from text token sequences in LibriSpeech dev-clean and dev-other \citep{panayotov2015librispeech}.

\textbf{Training} \method 
is trained on 8 NVIDIA A100 GPUs with DeepSpeed ZeRO-2 \citep{rasley2020deepspeed}. We utilize a cosine annealing learning rate scheduler with a peak learning rate of 1.5e-4, training the model with a batch size of 256 for 4 epochs. The maximum sequence length during training is 1200. For decoding, we use a temperature of 0.7, Top-$k$ sampling with $k=10$, and Top-$p$ sampling with $p=0.8$.

%% file: Tables/data.tex
\begin{table}[h]
\begin{center}
\resizebox{\textwidth}{18mm}{
\begin{tabular}{lll}
\toprule
\textbf{Task}&\textbf{Dataset}&\textbf{Items}\\
\midrule
ASR \& TTS & Multilingual Librispeech\citep{pratap2020mls}&20k hours\\
\midrule
Text QA & Guanaco\_belle\_merge\_v1.0 \citep{leng2023chinese-vicuna}&350k\\
\midrule
\multirow{2}{*}{Multi-turn Speech QA}& Moss-002-sft-data \citep{sun2024moss}&81k (242k turns)\\
&Coqa \citep{reddy2019coqa}&7k (103k turns)\\
\midrule
\multirow{2}{*}{Single-turn Speech QA}& SquAD\citep{rajpurkar2016squad}&87k\\
&Spoken-alpaca-gpt4 \citep{GSQA2024spoken}&54k\\
\bottomrule
\end{tabular}}
\caption{An overview of Datasets for training \method.}
\label{tab:data}
\end{center}
\end{table}

%% file: Sections/42_baselines.tex
We employed two baseline models based on SpeechGPT \citep{zhang2023speechgpt}, which share the same model weights but differ in their decoding prompts: (1) SpeechGPT, which processes only speech instructions, and (2) SpeechGPT-ASR, which processes both speech instructions and their corresponding transcriptions generated by Whisper en-medium \citep{radford2023robust}. In addition to the baselines above, we trained a model under the same training settings as \method, but employing the "reduced" strategies outlined in Section \ref{sec:33_GroupFormer}. This model, named IntrinsicVoice (Reduce), collapses consecutive sequences of identical units into a single unit, producing a sequence of unique discrete units.

%% file: Sections/43_evaluation.tex
We evaluate the cross-modal instruction-following in multi-turn dialogue scenarios across four tasks: \emph{speech-to-speech instruction-following} (S2SIF), \emph{text-to-speech instruction-following} (T2SIF), \emph{text-to-text instruction-following} (T2TIF), \emph{speech-to-text instruction-following} (S2TIF).

\textbf{Data} We construct our evaluation set,  IntrinsicVoice\_Eval, by randomly selecting 100 samples from the AlpacaEval dataset\footnote{https://github.com/tatsu-lab/alpaca\_eval}.
We then use CosyVoice to convert the text into speech. Each sample forms a quadruplet consisting of a speech instruction, text instruction, text response, and speech response, which we designate as ground truth.

\textbf{ChatGPT Score} Following \citet{zhang2023speechgpt}, we utilize GPT-4o to assess the content quality of response. For tasks involving speech, we employ Whisper-large-v3 \citep{radford2023robust} to transcribe the speech into text, which is subsequently evaluated. The detailed prompt used for evaluation is available in the Appendix \ref{sec:71_gpt_score}.

\textbf{Speech Quality} Following \citet{fang2024llama}, we utilize a Mean Opinion Score (MOS) prediction model, UTMOS\footnote{https://github.com/tarepan/SpeechMOS} \citep{saeki2022utmos}, to evaluate the quality and naturalness of the generated speech, referring to this metric as the UTMOS score.

\textbf{Response Latency} Latency is defined as the time delay between the end of the user’s utterance and the system’s initial speech response. We simulated latency for each sample in the evaluation set using a single NVIDIA A100 GPU and reported the median latency values.

%% file: Sections/44_main_results.tex
\input{Tables/results}

\textbf{Content} As shown in Table \ref{tab:result}, \method (Group) with TPS of 5 significantly outperformed \method (Reduce), which averaged 19.16 TPS, in both the S2SIF and T2SIF tasks regarding content quality. This result underscores the significant advantages of minimizing the gap between text token sequence length and speech token sequence length, thereby enhancing the transfer of textual capabilities to the speech domain. Furthermore, SpeechGPT+ASR demonstrated a notable performance advantage over SpeechGPT, indicating that the response content quality of SpeechGPT is highly dependent on the accuracy of the transcription provided for the speech instructions. In contrast, \method outperformed both SpeechGPT on the S2SIF and S2TIF tasks, showcasing its superior intrinsic speech understanding and modeling capabilities.

\textbf{Speech Quality} As shown in Table \ref{tab:result}, \method (Group) achieved UTMOS scores comparable to those of \method (Reduce), SpeechGPT, and SpeechGPT+ASR, indicating that the GroupFormer architecture enables the model to maintain high audio quality even at very low TPS.

\textbf{Latency}  \method achieved a latency of less than 100 ms, which is only one-tenth that of the SpeechGPT-based model, enabling real-time voice interaction. This experimental result highlights the significant advantage of generating speech responses directly from speech instructions, as opposed to relying on text autoregressive generation, in terms of real-time performance.

%% file: Tables/results.tex
\begin{table}[h]
\begin{center}
\resizebox{\textwidth}{20mm}{
\begin{tabular}{l|c|cccc|cc|c}
\toprule
&\textbf{TPS}&\multicolumn{4}{c|}{\textbf{ChatGPT Score} $\uparrow$}&\multicolumn{2}{c|}{\textbf{Latency (s)} $\downarrow$}&\textbf{UTMOS} $\uparrow$\\
\textbf{Methods} && S2SIF&S2TIF&T2SIF&T2TIF& S2SIF&T2SIF&\\
\midrule
Ground Truth&-&3.81 & 4.11 & 3.81 &4.11 &- &-&4.06\\
\midrule
\multicolumn{9}{l}{\emph{Baselines}}\\
SpeechGPT&-& 2.83 &3.04 & 3.41&3.72  &1.21&0.91 &3.73\\
SpeechGPT + ASR&-& 3.23& 3.31 &-  & -&1.07 &-&3.79\\

\midrule
\method (Group)&5 & 3.05 &3.22 &3.39 &3.87 &0.09 &0.09&3.75\\
\method (Reduce)&19.16&2.07 &2.97 & 2.23 &3.83 &0.10&0.10&3.81\\

\bottomrule
\end{tabular}}
\caption{Main Results of \method. TPS denotes tokens-per-second. S2SIF refers to speech-to-speech instruction-following, S2TIF is speech-to-text instruction-following, T2SIF denotes text-to-speech instruction-following and T2TIF represents text-to-text instruction-following. ChatGPT score is obtained through GPT-4o evaluation. Latency is the time delay between the end of the user’s utterance and the system’s initial speech response. UTMOS score is obtained by SpeechMOS. IntrinsicVoice (Reduce) is the model uses the "reduce" strategy to reduce the length of speech token sequences.}
\label{tab:result}
\end{center}
\end{table}

%% file: Sections/50_case_study.tex
Figure \ref{fig:case} illustrates a multi-turn speech QA example using \method. The results demonstrate that \method can understand speech instructions and respond appropriately during multi-turn interactions with humans.

\begin{figure*}[h]
    \centering
    \begin{tcolorbox}[width=1\textwidth]
\texttt{<|im\_start|>} system

You are a helpful voice-chat assistant\texttt{<|im\_end|>}

\texttt{<|im\_start|>} user

\texttt{<sosp>}\texttt{<speech>}...\texttt{<speech>}\texttt{<eosp>}\texttt{<|im\_end|>} 

\textcolor{blue}{\emph{Transcription:} Where is the capital of China}

\texttt{<|im\_start|>}assistant

\texttt{<sosp>}\texttt{<speech>}...\texttt{<speech>}\texttt{<eosp>}\texttt{<|im\_end|>} 

\textcolor{blue}{\emph{Transcription:} The capital of China is Beijing. Beijing is located in the northern part of China and is the political and economic center of the country.}

\texttt{<|im\_start|>}user

\texttt{<sosp>}\texttt{<speech>}...\texttt{<speech>}\texttt{<eosp>}\texttt{<|im\_end|>}

\textcolor{blue}{\emph{Transcription:} Can you recommend some attractions there}

\texttt{<|im\_start|>}assistant

\texttt{<sosp>}\texttt{<speech>}...\texttt{<speech>}\texttt{<eosp>}\texttt{<|im\_end|>}

\textcolor{blue}{\emph{Transcription:} Certainly, here are some must visit places:\\1. The Forbidden City - This vast palace complex was the imperial palace during the Ming and Qing dynasties. It's now a museum.\\
 2. The Great Wall of China - While parts of it are outside Beijing proper, sections like Badaling and Jinshanling offer breathtaking views and historical insights into one of the world's most iconic structures. \\3. Summer Palace - A beautiful imperial garden that features lakes, pavilions, temples, and ancient architecture, it's a serene escape from the city's hustle and bustle. \\4. Tiananmen Square - Not only a political and cultural center but also a significant historical site, Tiananmen Square is where many major events in Chinese history have taken place.
}

\end{tcolorbox}
    \caption{Case of multi-turn QA result}
    \label{fig:case}
\end{figure*}

%% file: Sections/60_conclusion.tex
In this paper, we proposed \method, an LLM with intrinsic real-time voice interaction capability. We introduced the GroupFormer architecture to reduce speech token sequences to lengths comparable to text token sequences while generating high-quality audio. This approach significantly mitigates the length disparity between speech and text, accelerates inference, and alleviates challenges associated with long-text modeling. Additionally, we constructed a multi-turn speech-to-speech dialogue dataset named \method-500k, which includes nearly 500,000 turns of speech-to-speech dialogues, and implemented a cross-modality training strategy to enhance semantic alignment between speech and text. Experimental results indicate that compressing speech token sequences to lengths similar to those of text token sequences effectively address ses the modality differences between speech and text, facilitating the transfer of pretrained LLM textual capabilities to the speech modality. Furthermore, \method achieves promising results in real-time multi-turn voice interaction scenarios.

%% file: Sections/70_appendix.tex
\subsection{ChatGPT Score Evaluation Prompt}
\input{Sections/71_gpt_eval_template}
\label{sec:71_gpt_score}

%% file: Sections/71_gpt_eval_template.tex
\begin{figure*}[ht]
    \centering
    \begin{tcolorbox}[width=1\textwidth]
You are evaluating a response that has been submitted for an instruction, using a specific set of standards. Below is the data:

***

[Instruction]: \textcolor{red}{\emph{(Instruction Transcription)}}

***

[Response]: \textcolor{red}{\emph{{(Response Transcription)}}}

***     

[Criterion]: helpfulness:

"1": "Not helpful - The generated text is completely irrelevant, unclear, or incomplete. It does not provide any useful information to the user."

"2": "Somewhat helpful - The generated text has some relevance to the user’s question, but it may
be unclear or incomplete. It provides only partial information, or the information provided may not be useful for the user’s needs."

"3": "Moderately helpful - The generated text is relevant to the user’s question, and it provides a
clear and complete answer. However, it may lack detail or explanation that would be helpful for
the user."

"4": "Helpful - The generated text is quite relevant to the user’s question, and it provides a clear,
complete, and detailed answer. It offers additional information or explanations that are useful
for the user. However, some of the points of the response are somewhat repetitive or could be
combined for greater clarity and concision"

"5": "Very helpful - The generated text is highly relevant to the user’s question, and it provides a
clear, complete, and detailed answer. It offers additional information, explanations, or analogies
that are not only useful but also insightful and valuable to the user. However, the structured of the
response is not well-organized and there is no clear progression or logical sequence of different
points in the response."

***

[END DATA]

Does the response meet the criterion? You should only write out your score in this format: "My
score is: "

\end{tcolorbox}
    \label{fig:gpt}
\end{figure*}